\documentclass[twocolumn]{revtex4-1}

\usepackage[T1]{fontenc}					% Zeichensatzkodierung
\usepackage[latin1]{inputenc} 				% Umlaute

\usepackage{graphicx}	 
\usepackage{tabularx}
\usepackage{amsmath}
\usepackage{amssymb}
\usepackage{mathrsfs}   
\usepackage{braket}
\usepackage{amsfonts}
\usepackage{amsthm}
\usepackage{dsfont}
\usepackage{mathtools}
\usepackage{todonotes}
\usepackage{enumerate}
\usepackage{color}   
\usepackage{subfig} 
\usepackage[unicode=true,pdfusetitle,
 bookmarks=true,bookmarksnumbered=false,bookmarksopen=false,
 breaklinks=true,pdfborder={0 0 0},backref=false,colorlinks=true]{hyperref}
\hypersetup{linkcolor=blue,citecolor=blue,urlcolor=blue}
\newcommand{\R}{\mathbb{R}}   
   
\newcommand{\1}{\mathds{1}}   

\newcommand{\vv}[1]{\boldsymbol{#1}}
\newcommand{\rot}{\nabla \wedge}

\newcommand{\Abs}[1]{\left| #1\right|}
\newcommand{\norm}[1]{| #1|}

\def\XXint#1#2#3{{\setbox0=\hbox{$#1{#2#3}{\int}$}
\vcenter{\hbox{$#2#3$}}\kern-.5\wd0}}

\begin{document}

\allowdisplaybreaks

\thispagestyle{empty}
    
\title{On the initial value formulation of classical electrodynamics}

\author{Dirk - Andr\'e Deckert}
\email{deckert@math.lmu.de}
\affiliation{
  Mathematisches Institut der Ludwig-Maximilians-Universit\"at M\"unchen\\
  Theresienstrasse 39, 80333 M\"unchen, Germany\\
}
\author{Vera Hartenstein}
\email{hartenstein@math.lmu.de}
\affiliation{
  Mathematisches Institut der Ludwig-Maximilians-Universit\"at M\"unchen\\
  Theresienstrasse  39, 80333 M\"unchen, Germany\\
}

\begin{abstract}
    We describe a seemingly unnoticed feature of the text-book Maxwell-Lorentz
    system of classical electrodynamics which challenges its formulation in
    terms of an initial value problem.  For point-charges, even after
    appropriate renormalization, we demonstrate that most of the generic
    initial data evolves to develop singularities in the electromagnetic fields
    along the light cones of the initial charge positions. We provide explicit
    formulas for the corresponding fields, demonstrate how this phenomenon
    renders the initial value problem ill-posed, and show how such bad initial
    data can be ruled out by extra conditions in addition to the Maxwell
    constraints.  These extra conditions, however, require knowledge of the
    history of the solution and, as we discuss, effectively turn the
    Maxwell-Lorentz system into a system of delay equations much like the
    Fokker-Schwarzschild-Tetrode equations. For extended charges such singular
    light fronts persist in a smoothened form and, as we argue, yield
    physically doubtful solutions. Our results also apply to some extent to
    expectation values of field operators in quantum field theory.  
\end{abstract}

\maketitle
  
\section{Introduction}
\label{sec:Intro}
In classical electrodynamics, the dynamics of $N$ charges and their
corresponding electromagnetic fields is governed by the Lorentz equations
\begin{align}
    \label{eq:lorentz}
    &\frac{d}{dt}\begin{pmatrix}
        \vv q_{i,t} \\ \vv p_{i,t}
    \end{pmatrix} 
    =  
    \begin{pmatrix}
        \vv v_{i,t} := \vv v(\vv p_{i,t})=\frac{\vv p_{i,t}}{\sqrt{\vv
            p_{i,t}^2+ m^2}} \\
            \sum_{j =1}^N e_{ij} \vv L_{ij,t} \\
        \end{pmatrix},
        \\
        \label{eq:lorentz_b}
        &\vv L_{ij,t}:=\int d^3x \, \rho( \vv x- \vv q_{i,t})
        [\vv E_{j,t} ( \vv x) + \vv v_{i,t} \wedge \vv B_{j,t}(\vv x)],
\end{align}
the Maxwell equations
\begin{align}
    \label{eq:maxwell}
    \partial_t \begin{pmatrix}
        \vv E_{i,t} \\ \vv B_{i,t}
    \end{pmatrix} 
    =  
    \begin{pmatrix}
        \nabla \wedge \vv B_{i,t}
        - 4 \pi \, \vv v_{i,t} \rho (\cdot - \vv q_{i,t})\\
        -\nabla \wedge \vv E_{i,t}
    \end{pmatrix},
\end{align}
and the Maxwell constraints
\begin{align}
    \label{eq:constraints}
    \nabla\cdot \vv E_{i,0} = 4\pi \, \rho(\cdot- \vv q_{i,0})
    \qquad
    \text{and}
    \qquad
    \nabla\cdot \vv B_{i,0} = 0,
\end{align}
for $i=1,\dots,N$.  In our notation, $\vv q_{i,t}, \vv p_{i,t} \in \R^3$ denote
the position and momentum of the $i$th charge at time $t\in\R$. For simplicity
we give all charges the same mass $m>0$ and rigid electric charge density
$\rho(\vv x)$ and use units such that the speed of light equals one and the
vacuum permittivity equals $(4\pi)^{-1}$. Note that by virtue of
\eqref{eq:maxwell}, the constraints \eqref{eq:constraints} at $t=0$ imply that
they hold for all times $t$.

Contrary to the text-book presentation, see, e.g., \cite{jackson1998,Roh07}, in
which one employs only one total electric and magnetic field, it will be
convenient for our discussion to associate with each charge $i$ an individual
electric and magnetic field $\vv F_{i,t}=(\vv E_{i,t}, \vv B_{i,t})$. Thanks to
the linearity of the Maxwell equations in the field degrees of freedom, the
equations of motion \eqref{eq:lorentz}-\eqref{eq:constraints} coincide with the
one given in text-books when setting $e_{ij}=1$. Other choices of $e_{ij}$ allow
to switch on or off the interaction of the $j$-th field on the $i$-th charge. 

For arbitrary $e_{ij}$ and smooth and compactly supported $\rho: \R^3 \to \R$ it
has been proven that the coupled system of equations
\eqref{eq:lorentz}-\eqref{eq:constraints} has a well-posed initial problem for
any initial data $(\vv q_{i,0},\vv p_{i,0},\vv F_{i,0})_{1\leq i\leq N}$ with
reasonably regular fields $\vv F_{i,0}$ fulfilling the constraints
\eqref{eq:constraints}; see \cite{KS00,BD01,BDD10}. Spinning charges were
discussed in \cite{AK01} and the semi-relativistic system was considered in
\cite{Falconi2013}.  Very early, however, it was
observed, e.g., in \cite{Abr03}, that replacing the charge density $\rho$ by a
Dirac delta distribution $\delta^3$ (for simplicity,
setting the total electric charge equal one) renders the self-interaction summand $\vv
L_{ii,t}$ on the right-hand side of the Lorentz equation \eqref{eq:lorentz},
and thereby, also the coupled system of equations
\eqref{eq:lorentz}-\eqref{eq:constraints}, ill-defined.  The reason for this is
that, in the point-charge case $\rho=\delta^3$, the Maxwell fields $\vv
F_{i,t}$ are not entirely smooth anymore but have a second order pole at $\vv
q_{i,t}$ which is  exactly where they would have to be evaluated in $\vv
L_{ii,t}$. In order to distinguish the case of general $\rho$ from the
point-charge case of $\rho=\delta^3$, we use the convention that lower-case
fields $\vv f_{i,t}$ solve the equations
\eqref{eq:maxwell}-\eqref{eq:constraints} for $\rho=\delta^3$, which then
implies the relation $\vv F_{i,t}=\rho*\vv f_{i,t}+ \vv F_{i,t}^0$, where $*$ denotes the
convolution and $\vv F_{i,t}^0$ is a solution to the free Maxwell equations,
i.e. \eqref{eq:maxwell}-\eqref{eq:constraints} for $\rho =0$.  To see the
divergent behavior of $\vv f_{i,t}$, thanks to the linearity, it suffices to
regard a special solution to \eqref{eq:maxwell}-\eqref{eq:constraints} for a
fixed charge trajectory $(\vv q_i,\vv p_i):t\mapsto (\vv q_{i,t}, \vv
p_{i,t})$. In the following we drop the index $i$ to keep the notation slim.
Two well-known solutions of \eqref{eq:maxwell}-\eqref{eq:constraints} are the
advanced and retarded Li\'enard-Wiechert fields $\vv f^\pm_t[\vv q,\vv p]=(\vv
e^\pm_t,\vv b^\pm_t)$, where the square bracket notation emphasizes the functional dependence
on the charge trajectory $(\vv q,\vv p)$. They are given by
\begin{align}
    \vv e^{\pm}_t(\vv x)
    &:=
    \frac{({\vv n}\pm\vv v)(1-\vv v^2)}
    {|\vv x - \vv q|^2(1\pm {\vv n}\cdot\vv v)^3}
    +
    \frac{{\vv n}\wedge[({\vv n}\pm\vv v)\wedge \vv a]}{|\vv x
    -\vv q|(1\pm {\vv n}\cdot\vv v)^3}\bigg|^\pm,
    \nonumber
    \\ 
    \vv b^{\pm}_t(\vv x)
    &:=
    \mp \vv n^{\pm}\wedge\vv e^{\pm}_t(\vv x),
    \label{eq:lw}
\end{align}
where we have used the abbreviations
\begin{align}
    \arraycolsep=0pt
    \label{eq:lw-abbreviations} 
    \begin{array}{rlrlc} 
        \vv q^\pm
        &:= \vv q_{t^\pm}, 
        \qquad\mbox{}
        &\vv v^\pm
        &:= \vv v(\vv p_{t^\pm}),
        \qquad\mbox{}
        & 
        \vv a^\pm
        := \frac{d}{dt}{\vv v}(\vv p_{t})|_{t=t^{\pm}},\\ 
        \vv n^\pm 
        &:= 
        \frac{\vv x-\vv q^\pm}{|\vv x-\vv q^\pm|},
        &
        t^\pm &:= t \pm |\vv x-\vv q^\pm|;
    \end{array}
\end{align}
cf.  \cite{Roh07,spohn_dynamics_2008}. All other solutions $\vv f_t$ to
\eqref{eq:maxwell}-\eqref{eq:constraints} for the same trajectory $(\vv q,\vv
p)$ can then be represented as
\begin{align}
    \label{eq:convex}
    \vv f_t=\lambda \vv f^-_t[\vv q,\vv p]+(1-\lambda)\vv
    f^+_t[\vv q,\vv p]+\vv f^0_t
\end{align} 
for $\lambda\in[0,1]$, where $\vv f^0_t$ is a solution to the corresponding
homogeneous equations, i.e., \eqref{eq:maxwell}-\eqref{eq:constraints} for
$\rho=0$. For smooth $\vv f^0_t$, the explicit expressions in \eqref{eq:lw}
imply that all corresponding fields $\vv f_t$ are smooth on $\mathbb
R^3\setminus\{\vv q_t\}$ where they admit the discussed singular behavior that
renders the term $\vv L_{ii,t}$ in \eqref{eq:lorentz_b} ill-defined for
$\rho=\delta^3$.

To still make sense out of this ill-defined self-interaction, an informal mass
renormalization argument is usually employed, see \cite{Dir38}, which
effectively replaces the problematic term $\vv L_{ii,t}$ with the finite
Abraham-Lorentz-Dirac back reaction $\vv L^{\mathrm{ALD}}_{ii,t}$. In the
non-relativistic regime, the latter may be approximated by $\vv
L^{\mathrm{ALD}}_{ii,t}\approx\frac{2}{3}e^2 \dddot{\vv q}_{i,t}$, with $e$ denoting
the electric charge.  This procedure cures the original problem, however,
introduces a dynamical instability as for almost all but very special initial
accelerations, which now must be provided along with initial positions and
momenta, the corresponding charge trajectories approach the speed of light
exponentially fast. Nevertheless, it was shown that the subset of physically
sensible solutions can be well approximated in certain regimes by a dynamically
stable version that was suggested by Landau and Lifschitz; see \cite{spohn_dynamics_2008}.

After replacing the ill-defined term $\vv L_{ii,t}$ appropriately or simply
omitting it by setting $e_{ij}=1-\delta_{ij}$, which often can be justified as
its renormalized version is usually small (e.g., for small acceleration, jerk,
and electric charge), one might hope that there are no further obstacles in
arriving at a solution theory for the Maxwell-Lorentz system
\eqref{eq:lorentz}-\eqref{eq:constraints} in the point-charge limit $\rho\to
\delta^3$. A general proof of the well-posedness of the corresponding initial
value problem, however, is difficult and remains open. The first two
difficulties are obvious: 1) The charges must not collide, otherwise $|\vv x-\vv
q^\pm|^{-2}$ in \eqref{eq:lw} blows up; and 2) the charges must not approach the
speed of light too fast, otherwise the factors $(1\pm \vv n^\pm\cdot\vv
v^\pm)^{-3}$ in \eqref{eq:lw} may blow up.  Mathematically, difficulty 1) poses
a similarly delicate problem as in the $N$-particle problem of gravitation, only
now with the additional complication that the Coulomb potentials in
\eqref{eq:lw} are Lorentz-boosted and to be evaluated at \emph{delayed} or
\emph{advanced} times $t^\pm$ as given in \eqref{eq:lw-abbreviations}.
Difficulty 2) is due to the accumulation of the escaping fields along the
light cone and must be excluded with an a priori bound on the charge velocities.
When handled with care, it is reasonable to expect that at most only very few
initial values $(\vv q_{i,0},\vv p_{i,0},\vv f_{i,0})_{1\leq i\leq N}$ lead to
catastrophic events due to these two difficulties. However, there is a third difficulty which is more subtle and, to our knowledge,
has not received attention yet. Given a charge trajectory $(\vv q_{i},\vv p_{i})$, only rather special initial
fields $\vv f_{i,0}$ give rise to solutions $\vv f_{i,t}$ to
\eqref{eq:maxwell}-\eqref{eq:constraints} that are sufficiently regular outside
a neighborhood of $\vv q_{i,t}$ in order to be evaluated in the terms $\vv
L_{ji,t}$ in \eqref{eq:lorentz_b} for all times. Generic initial fields will
generate singular fronts in the fields traveling at the speed of light, and another
charge $j$ having velocities below the speed of light is bound to traverse such
fronts in finite time. \\ 

In Section~\ref{sec:steps} we explain the mathematical origin of this questionable artifact and discuss how solutions with singular light fronts can
be ruled out by appropriate restrictions on the initial values. In
Section~\ref{sec:ml-system}, for the point-charge case $\rho=\delta^3$, we give
necessary conditions for global existence of piecewise as well as globally
smooth solutions to the Maxwell-Lorentz system
\eqref{eq:lorentz}-\eqref{eq:constraints}. We discuss that this
point-charge phenomenon has a straight-forward analogue in quantum
field theory, and furthermore, implications on the case of extended charges
$\rho$. In the latter, the singular light fronts qualitatively persist in
$\vv F_{i,t}$, however, in a smoothened version as can be seen from the convolution relation $\vv F_{i,t}=\rho*\vv f_{i,t}+ \vv F_{i,t}^0$. Due to this additional smoothness, the singular
light fronts cause no trouble concerning the solution theory anymore.
Nevertheless, as illustrated in a quantitative example in the end of Section
~\ref{sec:admissible}, they can cause sharp, though smooth, steps on the length
scale of the diameter of $\rho$, and therefore, in principal observable
radiation. Moreover, in Section~\ref{sec:admissible}, we demonstrate that the
initial value problem is ill-posed when demanding smooth global solutions, as
the necessary restriction on the initial fields $\vv f_{i,0}$ requires
information about the, at $t=0$, unknown charge trajectories $(\vv q_i,\vv
p_i)$. We introduce a mathematical procedure for finding admissible initial
fields despite this fact. The latter, however, introduces an unwanted
arbitrariness which, as we suggest in Section~\ref{sec:conclusion}, can be
eliminated  by physical reasoning. The resulting restrictions on the
initial values naturally turn the equations of motion of the Maxwell-Lorentz
system \eqref{eq:lorentz}-\eqref{eq:constraints} into a class of delay
differential equations that include the Fokker-Schwarzschild-Tetrode equations
of motion  of Wheeler-Feynman electrodynamics
\cite{Fokker,Tetrode,Schwarzschild,WF45,WF49} and the Synge equations
\cite{Syn40} as prime examples.

\section{Singular light fronts in the electrodynamic fields}
\label{sec:steps}

Let us assume for a moment that, at least for a neighborhood around $t=0$, the
Maxwell-Lorentz system \eqref{eq:lorentz}-\eqref{eq:constraints} has as a
solution with actual charge trajectories $(\vv q_i,\vv p_i)$ and fields $\vv
F_i:t\mapsto\vv F_{i,t}=(\vv E_{i,t},\vv B_{i,t})$ for $i=1,\dots,N$.  The goal
in this section is to introduce explicit formulas for those fields $\vv F_i$,
depending on their corresponding trajectory $(\vv q_i,\vv p_i)$ and initial
field $\vv F_{i,0}$, in order to infer the properties of general solutions to
\eqref{eq:maxwell}-\eqref{eq:constraints}.  Since we can always retrieve from
the point-particle fields $\vv f_{i,t}$ the ones of the extended charges by
convolution, $\vv F_{i,t}=\rho*\vv f_{i,t} + \vv F^0_{i,t}$, we
will consider the point-particle case $\rho= \delta^3$ only.  Furthermore, we
drop the index $i$ in this entire section because all the computations hold for
any given charge $i$. An explicit expression for $\vv f_t$ solving
\eqref{eq:maxwell}-\eqref{eq:constraints} for trajectory $(\vv q,\vv p)$ and
initial field $\vv f_0$ can be found by recasting the Maxwell equations in an
integral form that reads
\begin{align}
    \vv f_t &= \vv f^{(1)}_t + \vv f^{(2)}_t,
    \label{eq:kirchhoff}
    \\
    \vv f_t^{(1)} 
    &
    := 
    \begin{pmatrix}
        \partial_t & \rot \\ -\rot & \partial_t
    \end{pmatrix}
    K_{t-t_0} * \vv f_0,
    \label{eq:f1}
    \\
    \vv f_t^{(2)} 
    &
    := 
    4\pi \int_{t_0}^t ds 
    \begin{pmatrix}
        -\nabla & -\partial_t \\ 0 & \rot
    \end{pmatrix}
    K_{t-s} * 
    \begin{pmatrix}
        \delta^3(\cdot - \vv q_{s}) \\ \vv v_{s} \delta^3(\cdot - \vv q_{s})
    \end{pmatrix}. 
    \label{eq:f2}
\end{align}
Again, the convolution is denoted by $*$, and furthermore, $K_t$ is the
propagator of the wave equation given by
\begin{equation}
\label{eq:greensfn}
    K_t := K_t^- - K_t^+ 
    \quad
    \text{for}
    \quad 
    K_t^{\pm} := \frac{\delta(|\cdot|
    \pm t)}{4\pi |\cdot|},
\end{equation}
where $K^\pm_t$ are the advanced and retarded Green's functions of the
d'Alembert operator.  See Appendix~\ref{sec:appendix} for details on the
derivation of \eqref{eq:kirchhoff} from Kirchhoff's formulas. Note
that by virtue of the Maxwell equations \eqref{eq:maxwell}, the Maxwell
constraint \eqref{eq:constraints} is preserved over time, which then also
holds in this integral form \eqref{eq:kirchhoff}.

\begin{figure}[ht]
    \begin{center}
        \includegraphics[width=\linewidth]{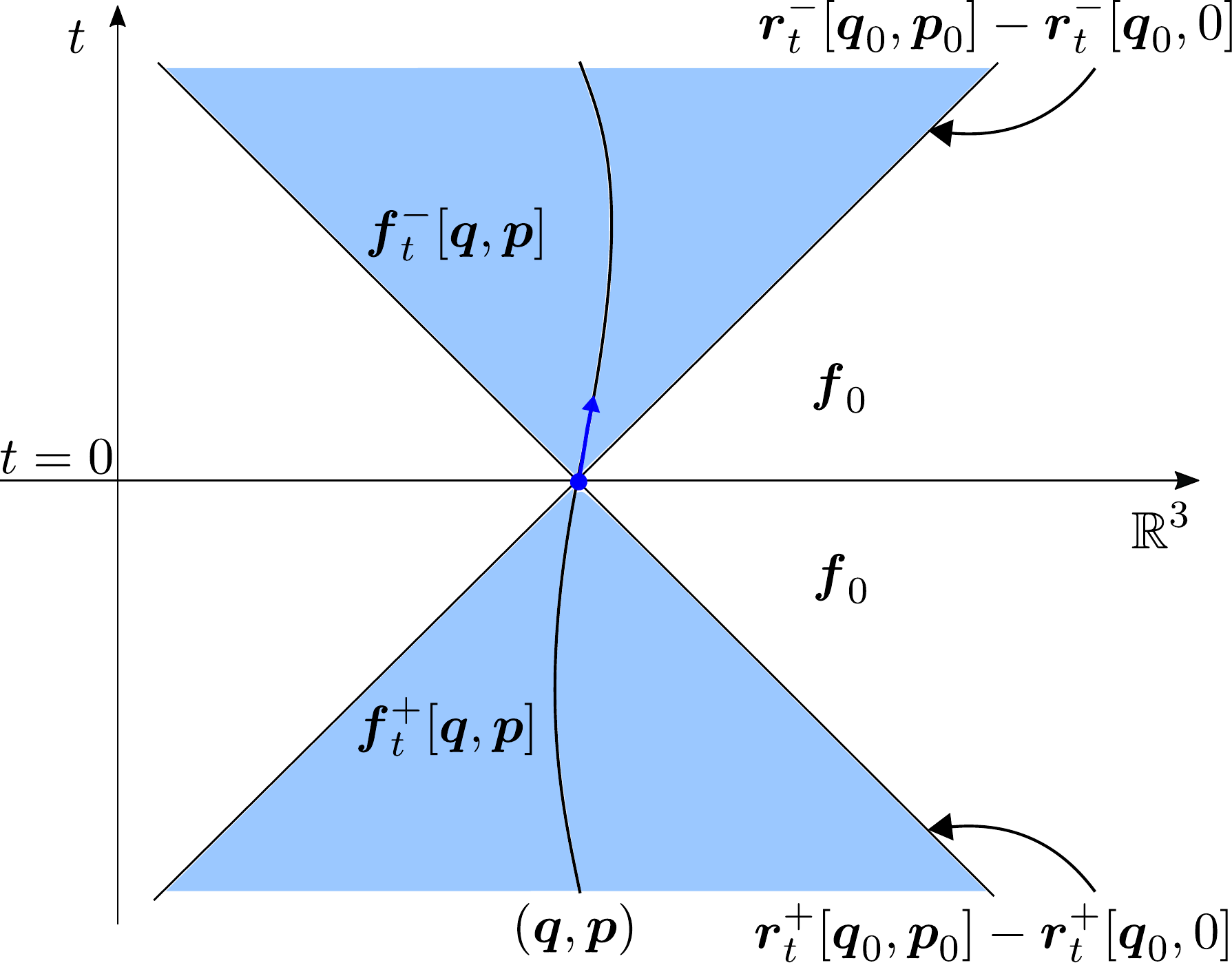}
    \end{center}
    \caption{\label{fig:coulomb} This figure illustrates the supports of the
    terms in \eqref{eq:coulomb_new}-\eqref{eq:coulomb_boundary} making up the
    solution $\vv f_t$ to the Maxwell equations
    \eqref{eq:maxwell}-\eqref{eq:constraints} for an initial Coulomb field $\vv
    f_0$ given in \eqref{eq:coulomb} and some fixed charge trajectory $(\vv
    q,\vv p)$. 
    Inside the light cone of $(0, \vv q_0)$ (blue area) the initial field $\vv f_0$ is displaced by the
    retarded/advanced Li\'enard-Wiechert field $\vv f_t^{- \sigma(t)}[\vv q, \vv p]$ in \eqref{eq:coulomb_new}. Outside that light cone (white area) the initial Coulomb field
    \eqref{eq:coulomb_old} persists. The distribution valued terms
    \eqref{eq:coulomb_boundary} are located only on the light cone.}
\end{figure}

Before we begin with an investigation of the properties of the general Maxwell field
\eqref{eq:kirchhoff}, it is illustrative to look at a simple example that shows
how singular light fronts arise. Considering the Maxwell constraint \eqref{eq:constraints}, one might think that
an obvious candidate for a good initial field $\vv f_0$ 
is given by the Coulomb field
\begin{align}
    \label{eq:coulomb}
    \vv f_0(\vv x) = (\vv e_0(\vv x),\vv b_0(\vv x)) = \left(\frac{\vv x-\vv
        q_{0}}{|\vv x-\vv q_{0}|^3},0\right).
\end{align}
Plugging the explicit
form of the initial field \eqref{eq:coulomb} into \eqref{eq:f1} and the actual
trajectory $(\vv q,\vv p)$ into \eqref{eq:f2} allows to compute
\eqref{eq:kirchhoff}. The corresponding solution $\vv f_t$ to
\eqref{eq:maxwell}-\eqref{eq:constraints} reads
\begin{align}
    \vv f_t 
    = 
    &
    \1_{B_{\Abs{t}}(\vv q_0)}\vv f_t^{-\sigma(t)}[\vv q, \vv p]
    \label{eq:coulomb_new}
    \\
    & 
    +\1_{B^c_{\Abs{t}}(\vv q_0)}\vv f_0
    \label{eq:coulomb_old}
    \\
    &
    + \vv r_t^{-\sigma(t)}[\vv q_0, \vv p_0],
    -\vv r_t^{-\sigma(t)}[{\vv q}_0,0],
    \label{eq:coulomb_boundary}
\end{align}
using
\begin{align}
    \vv r_t^{\pm }[\vv q_0, \vv p_0](\vv x):=  \frac{\delta(\Abs{t}- |\vv x - \vv q_0|) }{(1\pm \vv
        n_0 \cdot \vv v_0) |\vv x - \vv q_0|}   
        \begin{pmatrix}
            \vv n_0\pm
            \vv v_0 \\ - \vv n_0 \wedge \vv v_0
        \end{pmatrix},
        \label{eq:boundary}
\end{align}
together with
\begin{equation}
  \vv n_0 := \frac{\vv x - \vv q_0}{|\vv x - \vv q_0|}, \qquad
    \vv v_0 := \vv v (\vv p_0),
    \label{eq:abbreviations_0}
\end{equation} 
where, with slight abuse of the introduced square bracket notation, this
time the arguments in the square brackets in \eqref{eq:boundary} are not
functions but just position and momentum $\vv q_0, \vv
p_0\in\mathbb R^3$, respectively.  Furthermore, $\sigma(t)$ denotes the
sign of $t$, i.e., $\vv
f^{-\sigma(t)}_t$
stands for $\vv f^-_t$ if $t\geq 0$ and for $\vv f^+_t$ if $t<0$, and
$\1_{B_{\Abs t}(\vv q_0)}(\vv x)$ denotes the characteristic function being one for
$\vv x$ in the closed ball $B_{\Abs t}(\vv q_0)$ of radius ${\Abs t}$ around $\vv q_0$ and
zero for $\vv x$ in the open set $B_{\Abs t}^c(\vv q_0)=\R^3\setminus B_{\Abs t}(\vv q_0)$. For the details 
regarding the computation we refer the reader to the Appendix~\ref{sec:appendix}.  The
result shows that, according to \eqref{eq:coulomb_new}, inside the light cone
of space-time point $(t,\vv x)=(0,\vv q_0)$ the new advanced/retarded Li\'enard-Wiechert
field generated by the charge trajectory $(\vv q,\vv p)$ builds up as expected
while, according to \eqref{eq:coulomb_old}, in this region the initial Coulomb
field $\vv f_0$ given in \eqref{eq:coulomb} is displaced. The field $\vv f_0$
then persists only outside of that light cone. In addition, one finds two
distributions in \eqref{eq:coulomb_boundary} that depend on the Newtonian
initial data $(\vv q_0,\vv p_0)$ only and have support exclusively on the light cone. See Figure~\ref{fig:coulomb} for an illustration.

By inspecting \eqref{eq:coulomb_new}-\eqref{eq:coulomb_boundary} more closely
one thus finds that in general the Maxwell field $\vv f_t$ is not smooth on
$\R^3 \setminus\{\vv q_t\}$ although the initial field $\vv f_0$ in
\eqref{eq:coulomb} is. On the contrary, for most charge trajectories $(\vv
q,\vv p)$ the field will express singular fronts on the light cone of $(0,\vv q_0)$
because: 1) the distributions $\vv r^{-\sigma(t)}[\vv q_0,\vv
p_0]$ and $\vv r^{-\sigma(t)}[\vv q_0,0]$ in \eqref{eq:coulomb_boundary} cancel
only if $\vv p_0=0$; and 2) the
remaining terms \eqref{eq:coulomb_new} and \eqref{eq:coulomb_old} only connect
continuously on this light cone if at least the acceleration
$\lim_{t\downarrow 0} \ddot{\vv q}_t$ vanishes. Otherwise, the field $\vv
f_t$ will have a discontinuity there.  

At first sight this phenomenon may seem surprising. However, it has a rather
simple explanation. Morally, the initial Coulomb field $\vv f_0$ in
\eqref{eq:coulomb} corresponds to the field generated by a charge at rest at
position $\vv q_0$. More precisely, $\vv f_0$ is the retarded Li\'enard-Wiechert
field generated by an auxiliary charge trajectory $(\tilde{\vv q},\tilde{\vv
p})$ fulfilling $\tilde{\vv q}_0=\vv q_0$ and $\tilde{\vv p}_t=0$ for $t\leq 0$.
If the actual charge trajectory $(\vv q,\vv p)$ does not connect smoothly to
$(\tilde{\vv q},\tilde{\vv p})$ at $t=0$ but admits a kink, this sudden change
of acceleration will result in a radiation field traveling along the light cone of
$(0,\vv q_0)$. Should $\vv p_0$ be non-zero, an infinite acceleration is
necessary to change the momentum from $\tilde{\vv p}_0=0$ to $\vv p_0$, and the
corresponding radiation gives rise to the distributions \eqref{eq:boundary},
whereas a step in the acceleration merely causes a discontinuity on the
light cone. This simple example demonstrates that in order to prevent singular
light fronts, a compatibility condition between the initial field $\vv f_0$ and
the actual trajectory $(\vv q,\vv p)$  has to be met.

\begin{figure}[ht]
    \begin{center}
       \includegraphics[width=\linewidth]{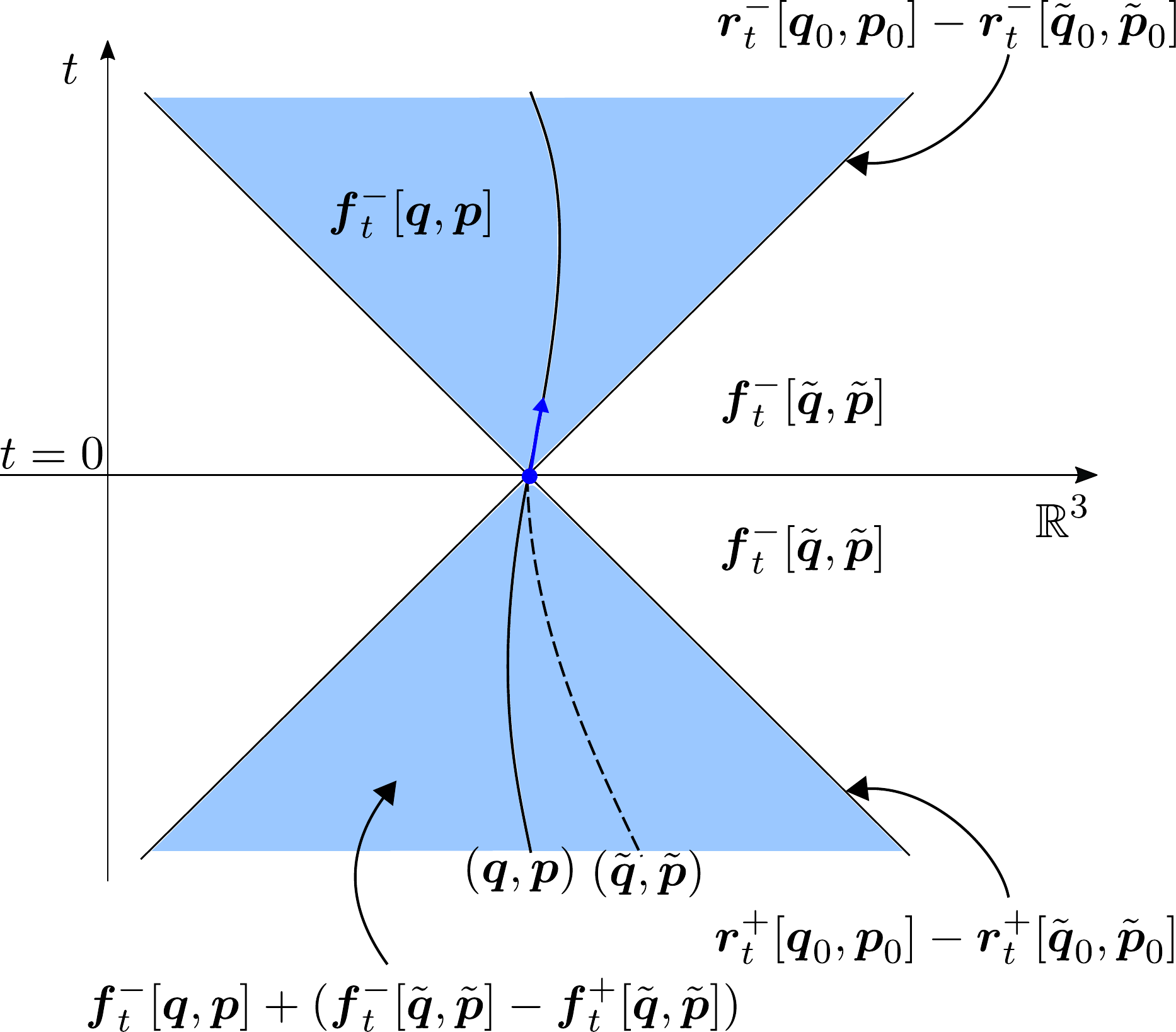} 
    \end{center}
    \caption{\label{fig:retarded}An illustration of the charge trajectories
    $(\vv q,\vv p)$ and $(\tilde{\vv q},\tilde{\vv p})$ as well as supports of
    the corresponding terms in \eqref{eq:f-new}-\eqref{eq:f-free} for the case
$\lambda=1$ and $\vv f_0^0=0$.}
\end{figure}

In the rest of this section, the objective is to identify a necessary
compatibility condition  for the general case. According to the general
splitting in \eqref{eq:convex}, also any relevant initial field $\vv
f_0$, obeying the Maxwell constraint \eqref{eq:constraints}, can be written in
the form 
\begin{equation}
    \label{eq:f_0}
    \vv f_{0} =  \lambda \vv f_{0}^-[\tilde{\vv q}, \tilde{\vv p}] +
    (1-\lambda) \vv f_{0}^+[\tilde{\vv q}, \tilde{\vv p}] + \vv
    f_{0}^{0},
\end{equation}
for some $\lambda\in[0,1]$, $\vv f^\pm_t[\tilde{\vv q},\tilde{\vv p}]$ being the
Li\'enard-Wiechert fields \eqref{eq:lw} generated by a smooth auxiliary charge
trajectory $(\tilde{\vv q},\tilde{\vv p})$ fulfilling $\tilde{\vv q}_0=\vv q_0$.
Note that, given any general initial field $\vv f_0$, equation \eqref{eq:f_0} is
merely a definition of $\vv f^0_0$ which must then be a homogeneous field, i.e.,
one fulfilling the Maxwell constraint \eqref{eq:constraints} for $\rho=0$. As
this free field $\vv f^0_t$ propagates independently of the charges,
nevertheless, influences them, it is reasonable to assume its initial value $\vv
f^0_0$ to be smooth (which implies $\vv f^0_t$ to be smooth) to avoid additional difficulties -- less regularity of $\vv
f^0_0$ and $(\tilde{\vv q},\tilde{\vv p})$ suffices, but this is not our focus
here.  Plugging the actual trajectory $(\vv q,\vv p)$ and the initial field $\vv
f_0$ in the form of \eqref{eq:f_0} into the explicit expressions
\eqref{eq:kirchhoff}-\eqref{eq:f2} above, one finds
\begin{align}
    \vv f_t 
    = 
    & 
    \1_{B_{\Abs{t}}(\vv q_0)}
    \vv f_t^{-\sigma(t)}[\vv q, \vv p] 
    \label{eq:f-new}
    \\
    & 
    +\1_{B_{\Abs{t}}(\vv q_0)}
    \lambda 
    \left(
            \vv f_t^{-}[\tilde{\vv q}, \tilde{\vv p}] 
            -
            \vv
            f_t^{-\sigma(t)}[\tilde{\vv q}, \tilde{\vv p}]
    \right)
    \label{eq:f-ret}
    \\
    &
    +
    \1_{B_{\Abs{t}}(\vv q_0)}
    (1- \lambda)  
    \Big(
        \vv f_t^{+}[\tilde{\vv q}, \tilde{\vv p}] 
        -
        \vv
        f_t^{-\sigma(t)}[\tilde{\vv q}, \tilde{\vv p}]
    \Big)
    \label{eq:f-adv}
    \\ 
    &
    +
    \1_{B_{\Abs{t}}^c(\vv q_0)}
    \Big(
            \lambda 
            \vv f_t^{-}[\tilde{\vv q}, \tilde{\vv p}] 
            +
            (1-\lambda)
            \vv f_t^{+}[\tilde{\vv q}, \tilde{\vv p}]
    \Big)
    \label{eq:f-old}
    \\
    &
    + \vv r_t^{-\sigma(t)}[\vv q_0, \vv p_0]-\vv r_t^{-\sigma(t)}[\tilde{\vv
    q}_0, \tilde{\vv p}_0]
    \label{eq:f-boundary}
    \\
    & 
    + \vv f_t^{0}.
    \label{eq:f-free}
\end{align}
The details are again given in the Appendix \ref{sec:appendix}. The
first three terms have support inside and on the light cone of $(0,\vv q_0)$. The
term \eqref{eq:f-new} describes the field that is generated by the actual charge
trajectory $(\vv q,\vv p)$ between time 0 and $t$, and terms
\eqref{eq:f-ret}-\eqref{eq:f-adv} describe how the initial advanced and retarded
Li\'enard-Wiechert fields encoded in \eqref{eq:f_0} are propagated inside the
light cone. Depending on the sign of $t$, one of the terms
\eqref{eq:f-ret}-\eqref{eq:f-adv} will vanish and the respective other will be
proportional to the difference $\sigma(t)(\vv f^+_t[\tilde{\vv q},\tilde{\vv
p}]-\vv f^-_t[\tilde{\vv q},\tilde{\vv p}])$, which according to Dirac
\cite{Dir38} can be interpreted as the radiation emitted or absorbed by the
auxiliary charge trajectory $(\tilde{\vv q},\tilde{\vv p})$ between time 0 and $t$.
Moreover, the term \eqref{eq:f-old} is the propagated remainder of the initial
retarded and advanced Li\'enard-Wiechert fields, and therefore, only has support
outside the light cone. The terms in \eqref{eq:f-boundary} are again the
distributions given in \eqref{eq:boundary} having support on the
light cone, and $\vv f_t^0$ in \eqref{eq:f-free} is simply the field $\vv f^0_0$
propagated from 0 to $t$ by the free Maxwell equations, i.e.,
\eqref{eq:maxwell}-\eqref{eq:constraints} for $\rho=0$. Note that $\vv f_t^0$ is
as regular as $\vv f_0^0$. See
Figure~\ref{fig:retarded} for an illustration of the trajectories and supports
of the terms \eqref{eq:f-new}-\eqref{eq:f-free}.
The solution $\vv f_t$ in \eqref{eq:f-new}-\eqref{eq:f-free} can be recast in a
more compact form
\begin{align}
    \vv f_t 
    =  
    &
    \1_{B_{\Abs{t}}(\vv q_0)}\left(\vv f_t^{-\sigma(t)}[\vv q, \vv p] -\vv
    f_t^{-\sigma(t)}[\tilde{\vv q}, \tilde{\vv p}] \right)
    \label{eq:short-new}
    \\ 
    &
    + \lambda \vv f_t^{-}[\tilde{\vv q}, \tilde{\vv p}] + (1- \lambda)  \vv
    f_t^{+}[\tilde{\vv q}, \tilde{\vv p}] 
    \label{eq:short-lw}
    \\ 
    &
    + \vv r_t^{-\sigma(t)}[\vv q_0, \vv p_0]-\vv r_t^{-\sigma(t)}[\tilde{\vv
    q}_0, \tilde{\vv p}_0]
    \label{eq:short-dist} 
    \\ 
    & 
    + \vv f_t^{0},
    \label{eq:short-free}
\end{align}
from which one can read off necessary compatibility conditions between
the initial field $\vv f_0$ and the charge trajectory $(\vv q,\vv p)$ that
prevent the development of singular light fronts:
\begin{itemize}
    \item[(C1)] The distributions \eqref{eq:short-dist} must cancel each other
       because neither \eqref{eq:short-new},
        \eqref{eq:short-lw}, nor \eqref{eq:short-free} contain Dirac delta
        distributions. This is the case if and only if 
        $(\tilde{\vv q}_0,\tilde{\vv p}_0)=(\vv q_0,\vv p_0)$, where $\tilde{\vv
        q}_0=\vv q_0$ was already assumed in order to fulfill the Maxwell constraint \eqref{eq:constraints}.
    \item[(C2)] Provided (C1) is fulfilled, the field $\vv f_t$  is continuous on
        $\mathbb R^3\setminus\{\vv q_t\}$ if and only if term
        \eqref{eq:short-new} vanishes on the light cone of
        $(0,\vv q_0)$. This can be seen as follows: By virtue of
        \eqref{eq:lw}, for all times $t$ the terms
        \eqref{eq:short-new}-\eqref{eq:short-lw} are smooth everywhere except 
        maybe on the light cone of $(0,\vv q_0)$ as well as the
        points $\vv q_t$ and $\tilde{\vv q}_t$. However, since these terms
        coincide with \eqref{eq:f-new}-\eqref{eq:f-old}, they must be smooth in
        $\tilde{\vv q}_t$ as \eqref{eq:f-ret} and \eqref{eq:f-adv} are free
        fields and \eqref{eq:f-old} has only support outside of the light cone
        of $(0,\vv q_0=\tilde{\vv q}_0)$. As the free field $\vv f^0_t$ is
        smooth, and by (C1) terms \eqref{eq:short-dist}, \eqref{eq:f-boundary}
        vanish, the field $\vv f_t$ is continuous on $\mathbb R^3\setminus\{\vv
        q_t\}$ if and only if \eqref{eq:short-new} vanishes on the
       light cone of $(0,\vv q_0)$. This is the case if and only
        if the accelerations $\ddot{\tilde{\vv q}}_t$ and $\ddot{\vv q}_t$
        coincide at $t=0$.  Furthermore, if and only if  all $l$-th derivatives
        of $\tilde{\vv q}_t$ and $\vv q_t$ for $l=1,\dots,k+2$ coincide at
        $t=0$, the field $\vv f_t$ has $k$ spatial derivatives on
        $\R^3\setminus\{\vv q_t\}$. Finally, if and only if  the trajectories
        $\tilde{\vv q}_t$ and $\vv q_t$ connect smoothly at time $t=0$, the
        field $\vv f_t$ is smooth on $\R^3\setminus\{\vv q_t\}$.
\end{itemize}

It was called to our attention that also in
\cite{Noja1998,Noja1999}, where a rigorous electrodynamic point-charge limit
was studied in the dipole approximation, a condition relating the initial
fields and initial momenta similar to (C1) was needed to ensure convergence.

\section{Implications on the Maxwell-Lorentz system}
\label{sec:ml-system}

In this chapter we discuss the implications of the observations made in
Section~\ref{sec:steps}  on the fully coupled system of Maxwell's and Lorentz's
equations \eqref{eq:lorentz}-\eqref{eq:constraints}. Our main interest, which
will be discussed first, lies in the case of $N \geq 2$ point-like charges, i.e.,
$\rho=\delta^3$, either with a properly renormalized self-interaction term $\vv
L_{ii,t}$ or without it, i.e., $e_{ij}=1-\delta_{ij}$.  The implications on the
Maxwell-Lorentz system for smooth extended charges $\rho$ are considered in the
end. 

\begin{figure}[ht]
    \begin{center}
        \includegraphics[width=\linewidth]{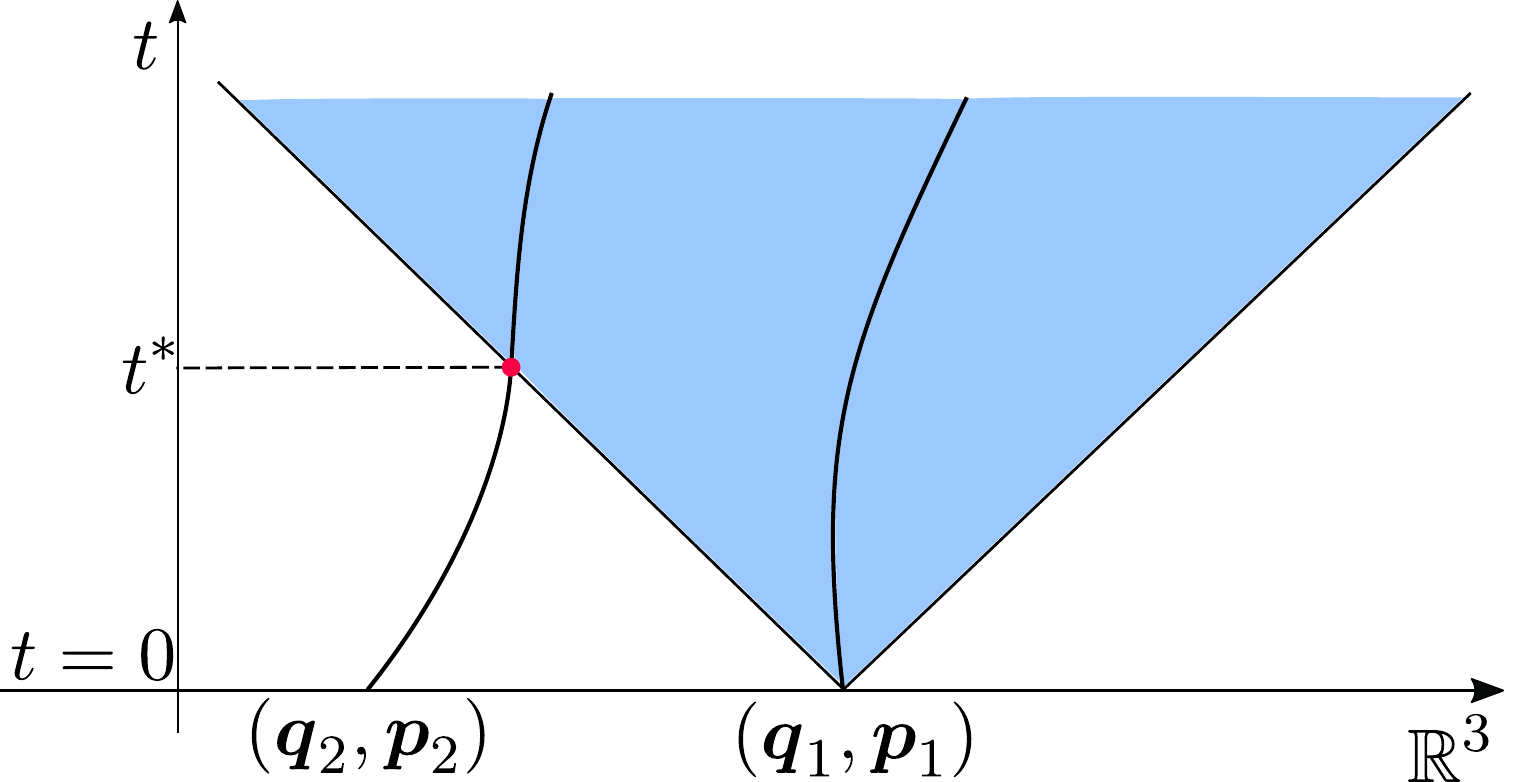}
    \end{center}
    \caption{\label{fig:3}Charge 2 on trajectory $(\vv q_2,\vv p_2)$  
    is bound to cross the light cone of the initial space-time
    point $(0,\vv q_{1,0})$ of charge 1 on trajectory $(\vv q_1,\vv p_1)$.}
\end{figure}

First and foremost, we observe that in a system of at least two charges, one
charge, say number 2, will inevitably cross the light cone of
the initial space-time point of another charge, say number 1, at a time $t^*$,
which is bounded from below by the minimal distance divided by speed of light;
see Figure~\ref{fig:3}.
Thus, at $t=t^*$ the Lorentz force \eqref{eq:lorentz} felt by charge 2 must
evaluate the field $\vv f_{1,t}$ at some point on the light cone
of $(0,\vv q_{1,0})$. Recall that for an initial field $\vv f_{i,0}$ of the form
\eqref{eq:f_0} with auxiliary charge trajectory $(\tilde{\vv q}_i,\tilde{\vv
p}_i)$, the propagated field $\vv f_{i,t}$ is given by
\eqref{eq:f-new}-\eqref{eq:f-free}. Should condition (C1) of
Section~\ref{sec:steps} not be satisfied, this evaluation is ill-defined because
of the presence of the distributions \eqref{eq:f-boundary}.  In this case, the
dynamics will cease to exist beyond the time instant $t^*$.  Hence, (C1) is a
necessary condition for global existence of solutions to the Maxwell-Lorentz
system \eqref{eq:lorentz}-\eqref{eq:constraints}.  Should condition (C1) hold
but not (C2), then the force on charge 2 will undergo a discontinuous jump when
traversing the light cone at time $t^*$.  Therefore, (C2) is a
necessary condition for having continuous or smooth solutions to the
Maxwell-Lorentz system \eqref{eq:lorentz}-\eqref{eq:constraints}.

The following two arguments illustrate that (C1) and (C2) are violated for
generic initial data $(\vv q_{i,0},\vv p_{i,0},\vv f_{i,0})_{1\leq i\leq N}$
obeying the Maxwell constraints \eqref{eq:constraints} only. Precisely, they show
that global existence is not stable under arbitrarily small perturbations of the
initial data.  For this purpose, let us assume that $(\vv q_i,\vv p_i,\vv
f_i)_{1\leq i\leq N}$ is a global solution to the Maxwell-Lorentz system
\eqref{eq:lorentz}-\eqref{eq:constraints} for some initial value $(\vv
q_{i,0},\vv p_{i,0},\vv f_{i,0})_{1\leq i\leq N}$ such that the initial fields
$\vv f_{i,0}$ are of the form \eqref{eq:f_0} for some smooth auxiliary
trajectory $(\tilde{\vv q}_i,\tilde{\vv p}_i)$ and some smooth initial free
field $\vv f^0_{i,0}$. Recall from our discussion in Section~\ref{sec:steps} that
any relevant initial field can be written in this form, and then, it automatically
fulfills the Maxwell constraint \eqref{eq:constraints}.\\

\begin{figure}[ht]
    \begin{center}
        \includegraphics[width=\linewidth]{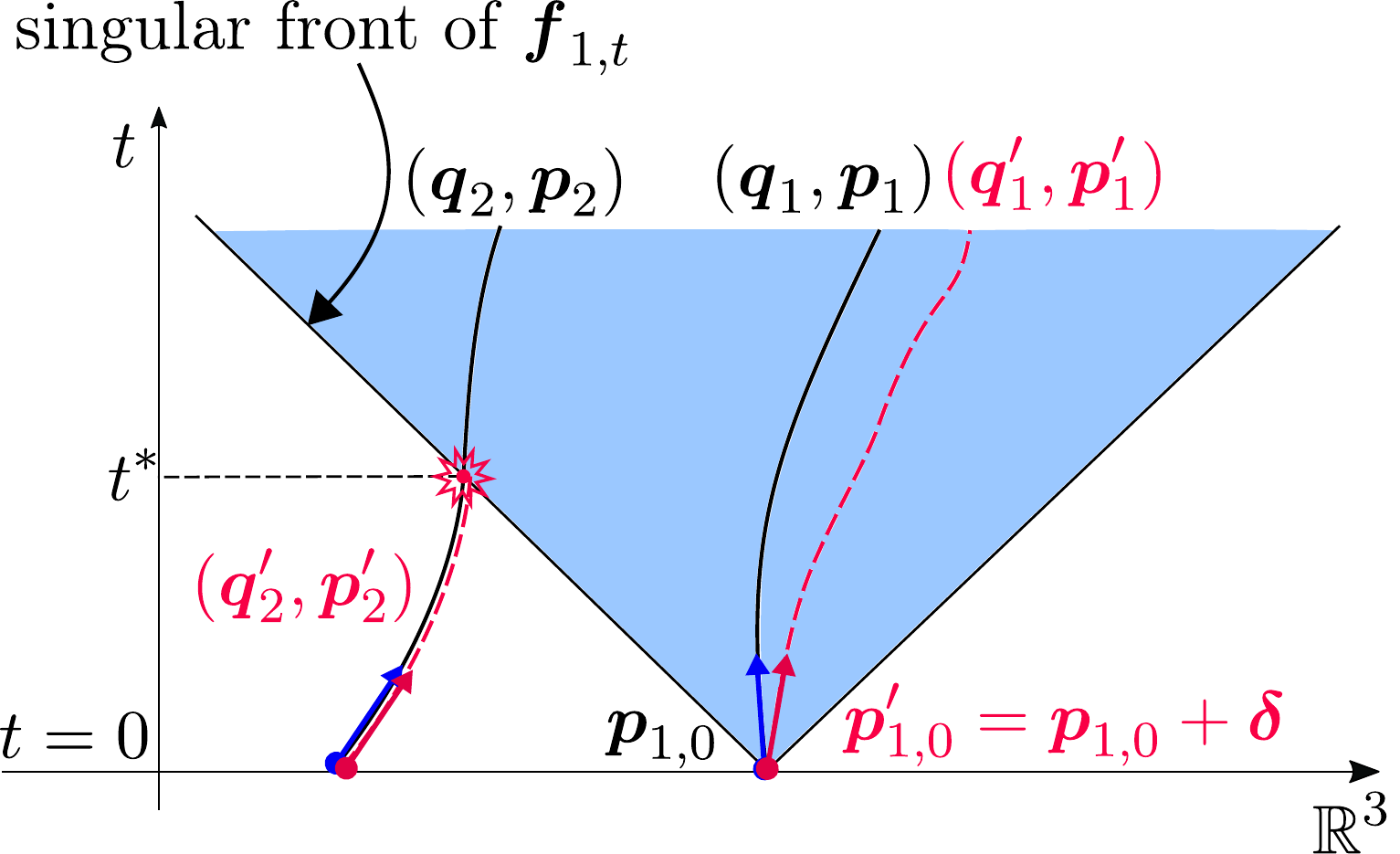}
    \end{center}
    \caption{\label{fig:A1} Perturbing the initial momentum of charge 1 by $\vv
    p_{1,0} \to \vv p_{1,0}'$ leads to a singular front supported on the light cone of $(0,\vv q_{1,0})$ and, thus, to a sudden stop of the
    dynamics at time $t^{*}$ when charge 2 touches the light cone.}
\end{figure}

\textbf{No-go argument (A1):}  By Maxwell constraints \eqref{eq:constraints} and necessary condition (C1) we have $(\tilde{\vv q}_{i,0},\tilde{\vv p}_{i,0})=(\vv
q_{i,0},\vv p_{i,0})$ for $i=1,\dots,N$. Then, perturbing the initial momentum of charge 1 by $\vv p_{1,0}\to\vv
p_{1,0}'=\vv p_{1,0}+\vv \delta$ for any vector $\vv \delta$ of arbitrarily small norm $|\vv \delta|>0$ leads to a corresponding local solution $(\vv q_i',\vv p_i',\vv
f_i')_{1\leq i\leq N}$ with $\vv f_{1,t}'$ taking the form of
\eqref{eq:f-new}-\eqref{eq:f-free}, whereas the contribution \eqref{eq:f-boundary}
equals the distribution $\vv r_t^{-\sigma(t)}[\vv q_0, \vv p_0+\vv \delta]-\vv
r_t^{-\sigma(t)}[{\vv q}_0, {\vv p}_0]$, which does not vanish.
In other words (C1) is violated, and $\vv f_{1,t}'$ manifests a singular light front with
support on the light cone of space-time point $(0,\vv q_{1,0})$,
as discussed in Section~\eqref{sec:steps}. By virtue of
\eqref{eq:lorentz}-\eqref{eq:constraints}, this perturbation in the initial
momentum propagates not faster than the speed of light. In particular, the perturbed
field $\vv f_{1,t}'$ of charge 1 and the perturbed trajectory $(\vv
q_{2,t}',\vv p_{2,t}')$ of charge 2 remain identical on $B^c_{|t|}(\vv q_{1,0})$
for $t\in\R$. In consequence, charge 2 is bound to touch the light cone of $(0,\vv q_{1,0})$ at the very same time $t^*$ as in the
unperturbed solution, only now the perturbed field $\vv f_{1,t}'$ contains a
singular light front consisting of distributions. In conclusion, the dynamics will cease
to exist beyond time $t^*$, as discussed above. The argument is depicted in
Figure~\ref{fig:A1}.

\begin{figure}[ht]
    \begin{center}
         \includegraphics[width=\linewidth]{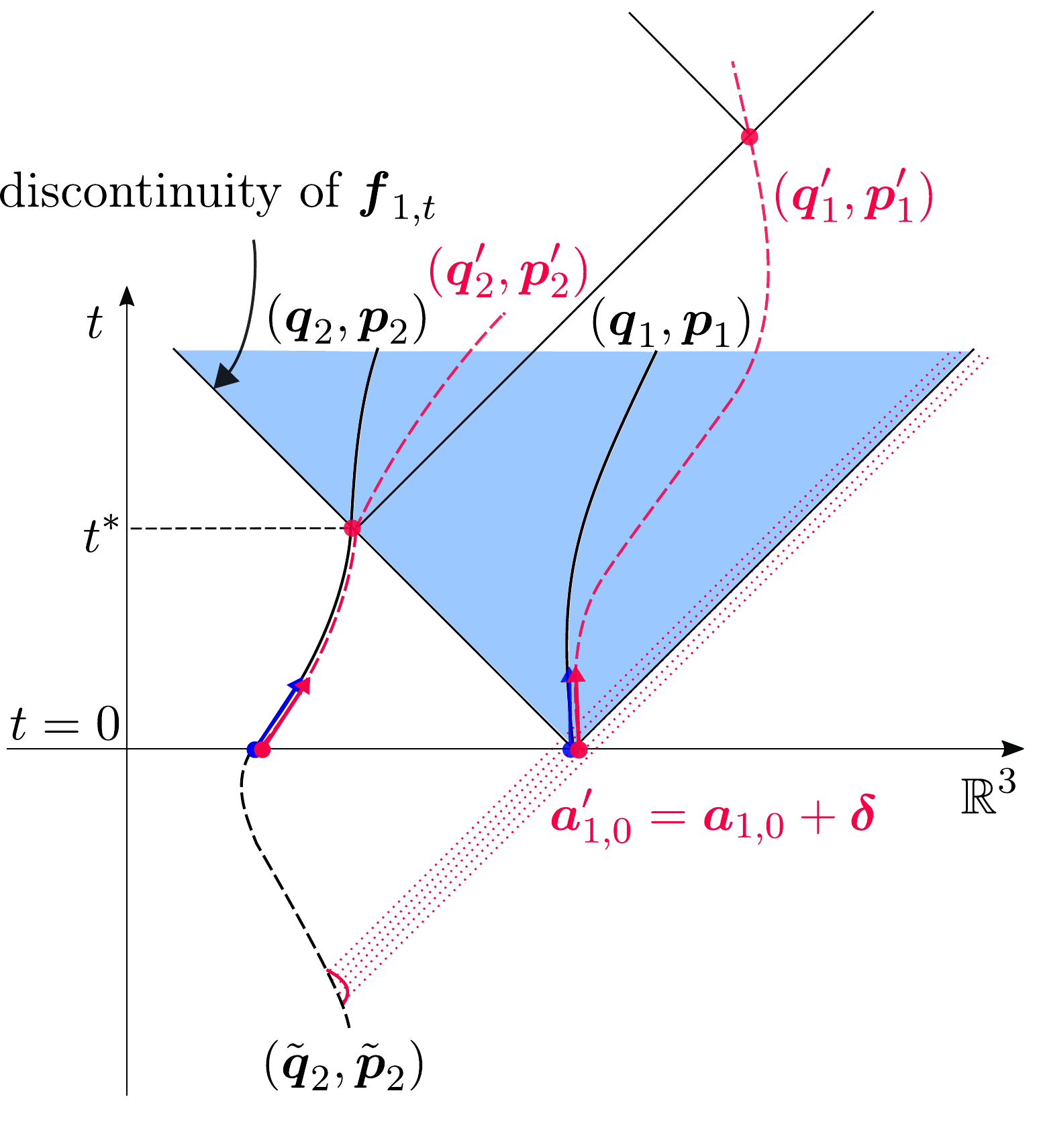}
    \end{center}
    \caption{\label{fig:A2} Perturbing the initial field of charge 1, $\vv
    f_{1,0}\to \vv f_{1,0}'$, by a small bump in $(\tilde{\vv q}_2,\tilde{\vv
    p}_2)$ at the corresponding retarded time, leads to a discontinuity of $\vv
    f_{1,t}$
    supported on the light cone of $(0,\vv q_{1,0})$, and thus,
    charge 2 experiences a sudden jump in acceleration at time $t^{*}$, which
    causes a discontinuity in $\vv f_{2,t}$.}
\end{figure}

\textbf{No-go argument (A2):} This time, let us assume the global solution is
also smooth and $\lambda>0$ (for $\lambda=0$ a similar argument can be found).
Due to condition (C2), $\ddot{\tilde{\vv q}}_{i,t}$ and $\ddot{\vv q}_{i,t}$
coincide at time $t=0$. Now, we perturb a little bit the trajectory $(\tilde{\vv
q}_2,\tilde{\vv p}_2)$ that defined the initial field $\vv f_{2,0}$ given in
\eqref{eq:f_0} in an arbitrarily small neighborhood of the retarded time
$t^-$ belonging to space-time point $(0,\vv q_{1,0})$. Due to
\eqref{eq:f-new}-\eqref{eq:f-free} this causes a small perturbation
$\vv f_{2,0}\to \vv f_{2,0}'$, and we tune this perturbation such that the
Lorentz force \eqref{eq:lorentz} on charge 1 at $t=0$ changes its value. In
consequence, the potential local solution $(\vv q_i',\vv p_i',\vv f_i')_{1\leq
i\leq N}$ corresponding to this perturbed initial data violates (C2) as the
accelerations $\ddot{\tilde{\vv q}}_{1,t}$ and $\ddot{\vv q}'_{1,t}$ do not
match anymore at $t=0$. As discussed in Section~\eqref{sec:steps}, this leads to a discontinuity on the light cone of $(0,\vv q_{1,0})$. However, by virtue of
\eqref{eq:lorentz}-\eqref{eq:constraints} the perturbed field $\vv f_{1,t}'$ of
charge 1 and the perturbed trajectory $(\vv q_{2,t}',\vv p_{2,t}')$ of charge 2
remain identical on $B^c_{|t|}(\vv q_{1,0})$ for $t\in\R$. Therefore,
charge 2 is bound to hit the light cone of $(0,\vv q_{1,0})$ at
the very same time $t^*$ as in the unperturbed solution. At this instant, due to the
discontinuity of $\vv f_{1,t}'$, the acceleration of charge 2 will undergo a likewise
discontinuous jump.  Hence, should the perturbed solution exist globally it can
only be piecewise smooth. Furthermore, the discontinuity in the acceleration of
charge 2 will give rise to a corresponding discontinuity in the field $\vv
f_{2,t}$ on the light cone of $(t^*,\vv q_{2,t^*})$, which
charge 1 is bound to cross eventually. By this mechanism, a whole network of
singular light fronts is developed. The argument is depicted in Figure~\ref{fig:A2}.\\

These two arguments indicate that the initial value problem of the
Maxwell-Lorentz system \eqref{eq:lorentz}-\eqref{eq:constraints} with
renormalized (or without) self-interaction term  is ill-posed for general
initial values $(\vv q_{i,0},\vv p_{i,0},\vv f_{i,0})_{1\leq i\leq N}$ only
fulfilling the Maxwell constraints \eqref{eq:constraints}: Even if a global
solution is found, only a small perturbation in the initial values suffices to
prevent either global existence, by (A1), or global smoothness, by (A2), of the
potential solution  corresponding to the perturbed initial values. 

One might tend to think that these are all problems connected to the
point-like nature of the charges, a concept that could even be considered 
questionable in the classical regime. 

Indeed, as discussed in the introduction, it is true that for the
Maxwell-Lorentz system of smoothly extended charges those mathematical problems
do not show up.  Nevertheless, the qualitative behavior of generation of
singular light fronts for initial conditions that violate (C1) or (C2) remains
the same. As the fields of the extended charges are of the form $\vv
F_{i,t}=\rho*\vv f_{i,t}+ \vv F_{i,t}^0$, the discussed singular fronts
are now only smeared out by the charge density $\rho$.  For $\rho$ supported on
the scale of the classical electron radius, i.e., $r_e\sim10^{-15}\text m$, the
singular fronts will still result in sharp -- though smooth -- steps in the
fields on the respective light cones. Other charges are bound to eventually
traverse such steps and will suddenly~--~on time scales of $r_e$ divided by
their respective speed~--~start or stop to radiate, thus, leading to
potentially observable though physically questionable phenomena. 

Furthermore, it is interesting to note that the singular light fronts persist
also in quantum field theory. This can readily be observed in the following toy
model in which a fixed source at $\vv q\in\mathbb R^3$ interacts with a
second-quantized and massless scalar field
\begin{align*}
    \varphi(t,\vv x) 
    = 
    \int d^3k \frac{(2\pi)^{-\frac32}}{\sqrt{2|\vv k|}}
    (a_{\vv k}e^{i\vv k\cdot\vv x-i\omega_{\vv k}t}+\operatorname{c.c.})
\end{align*}
by means of the interaction Hamiltonian $H_I(t) = g\varphi(t,\vv q)$, where
$g\in\mathbb R$ and $a_{\vv k}, a^*_{\vv k}$ are
the bosonic creation and annihilation operators fulfilling the 
CCR $[a_{\vv k},a_{\vv k'}^*]=\delta^3(\vv k-\vv k')$. Formally, $H_I(t)$ is
not well-defined without an ultraviolet cut-off such as $\rho$ but for the sake
of the argument it is sufficient to continue informally.
Let $U_I(t)$ be the time evolution generated by $H_I(t)$, then
for any initial unit Fock state $|\Psi\rangle$ we get
\begin{align*}
    \Box \, \langle \Psi | U_I(t)^* \varphi(t,\vv x)U_I(t) | \Psi\rangle = -g
    \delta^3(\vv x-\vv q).
\end{align*}
The expectation value of this scalar field can therefore be represented by
means of Kirchhoff's formulas as it was done for the Maxwell field in
Appendix~\ref{sec:appendix}. In the simplest case of an initial vacuum
$|\Psi\rangle=|0\rangle$ one finds
\begin{align*}
    & \langle0| U_I(t)^* \varphi(t,\vv x)U_I(t) | 0\rangle \\
    & =
    -g\int_0^t ds\, (K_{t-s}*\delta^3(\cdot-\vv q))(\vv x) 
    = -\frac{g}{4\pi} \frac{\1_{B_{\Abs{t}}(\vv q)}(\vv x)}{|\vv x-\vv q|}, 
\end{align*}
where the discontinuity on the light front shows up again as the field is build
up over time starting from an initial vacuum. This behavior only disappears for
special initial $|\Psi\rangle$, precisely, the ground state of this toy model,
which can be computed explicitly, plus smooth additional free fields. If we
further allow the charge $\vv q$ to move, very similar scenarios as discussed in
Section~\ref{sec:steps} can be constructed; but this shall not be our focus
here. \linebreak

In conclusion, for any choice of $\rho$, and be it for mathematical or physical
reasons, it seems desirable to restrict the space of initial values of the
Maxwell-Lorentz system \eqref{eq:lorentz}-\eqref{eq:constraints} beyond the
Maxwell constraints \eqref{eq:constraints}.

\section{Admissible Initial Values}
\label{sec:admissible}

If for a moment we also admit piecewise smooth solutions to the Maxwell-Lorentz
system \eqref{eq:lorentz}-\eqref{eq:constraints}, a sensible restriction on the
space of initial values can be taken from  condition (C1). If we require the
initial value $(\vv q_{i,0},\vv p_{i,0},\vv f_{i,0})_{1\leq i\leq N}$ to
comprise fields $\vv f_{i,0}$ of the form \eqref{eq:f_0} for piecewise smooth
auxiliary trajectories $(\tilde{\vv q}_i,\tilde{\vv p}_i)$ fulfilling
$(\tilde{\vv q}_{i,0},\tilde{\vv p}_{i,0})=({\vv q}_{i,0},{\vv p}_{i,0})$,
condition (C1) as well as the Maxwell constraints \eqref{eq:constraints} are
fulfilled by definition and there seems to be no further obstacle concerning
mathematical well-posedness of the respective initial value problem.

If, however, we demand smooth global solutions, we would also need to comply
with condition (C2). In order to do so we would have to know the derivatives of
the charge trajectories $(\vv q_i,\vv p_i)$ at initial time $t=0$.  But those
are unknown as they already require knowledge of a local solution in a
neighborhood of $t=0$.
Hence, there is no possibility to restrict the space of initial fields 
a priori in order to ensure well-posedness.

As a workaround one may consider the following approach:  Given initial data
$(\vv q_{i,0},\vv p_{i,0},\vv f_{i,0})_{1\leq i\leq N}$ fulfilling
\eqref{eq:constraints} and (C1), it is possible to compute the solution of the
Maxwell-Lorentz equations in a sufficiently small time interval $[0,\tau)$.
This can be done as the singular fronts live only on the light cones of the initial space-time points $(0,\vv q_{i,0})$ so that
$\tau$ only has to be chosen smaller than the smallest time $t^*$ when some
charge hits a singular front.  This preliminary local solution allows to
compute all derivatives of the charge trajectories $(\vv q_i,\vv p_i)$ at
$t=0$, and hence, it would allow to adapt the auxiliary trajectories
$(\tilde{\vv q}_i,\tilde{\vv p}_i)$ in a neighborhood of $t=0$ to connect
smoothly to $(\vv q_i,\vv p_i)$ such that (C2) is fulfilled. This
procedure changes the initial fields $\vv f_{i,0}\to\vv f_{i,0}'$ in a spatial
neighborhood around the initial positions $\vv q_{i,0}$. If self-interaction is
excluded, the adapted initial values $(\vv q_{i,0},\vv p_{i,0},\vv
f_{i,0}')_{1\leq i\leq N}$, however, fulfill the Maxwell constraints
\eqref{eq:constraints}, (C1), and (C2), and therefore, should not bare any
further obstacles concerning smooth global solutions. If
self-interaction is included, the above procedure would have to be iterated
until a fixed-point is found as the change in $\vv f'_{i,0}$ implies again a 
change in the initial acceleration of charge $i$. \\

Though mathematically sound, physically, this is a rather opaque procedure. It
is not anymore a formulation of classical electrodynamics in terms of an initial
value problem for \eqref{eq:lorentz}-\eqref{eq:constraints} but in terms of an
initial guess, that, first, has to be adapted in a quite arbitrary way before
a global solution can be inferred at all. \\

So what is overlooked when naively regarding the Maxwell-Lorentz system
\eqref{eq:lorentz}-\eqref{eq:constraints} as an initial value problem? Any
inhomogeneous solution $\vv f_{i,t}$ to the Maxwell equations
\eqref{eq:maxwell}-\eqref{eq:constraints} is of the form \eqref{eq:convex},
which implies that the entire history
of the charge trajectory $(\vv q_i,\vv p_i)$ is already encoded in the spatial
dependence of the field $\vv f_{i,t}$; recall the $t^\pm$ dependence in
\eqref{eq:lw}.  Now, if we set some initial field $\vv f_{i,0}$ by hand, for
which the Maxwell constraint \eqref{eq:constraints} only requires that we choose
it of the form \eqref{eq:f_0} with some auxiliary trajectory $(\tilde{\vv
q}_i,\tilde{\vv p}_i)$ fulfilling $\tilde{\vv q}_{i,0}=\vv q_{i,0}$, the Maxwell
time evolution is fooled to believe that the history of the charge trajectory is
given by $(\tilde{\vv q}_i,\tilde{\vv p}_i)$. But except for $\tilde{\vv
q}_{i,0}=\vv q_{i,0}$, the history of the auxiliary trajectory $(\tilde{\vv
q}_i,\tilde{\vv p}_i)$ may have nothing in common with the actual one $({\vv
q}_i,{\vv p}_i)$, which is to be computed. As a matter of fact, the Maxwell
equations propagate such an initial field $\vv f_{i,0}$ as if it was generated by
the auxiliary charge trajectory $(\tilde{\vv q}_i,\tilde{\vv p}_i)$ outside
the light cone of $(0,\vv q_{i,0})$ while, inside, a new
field is generated  according to the actual trajectory $(\vv q_i,\vv
p_i)$. It is therefore not surprising that the 
incompatibilities between the actual charge trajectories $(\vv q_i,\vv p_i)$ and
the initial fields $\vv f_{i,0}$ of the solution
\eqref{eq:lorentz}-\eqref{eq:constraints} discussed in Section~\ref{sec:steps}
occur during the dynamics and that any mismatch between the actual and auxiliary
charge trajectories in the sense of (C1) and (C2) expresses itself as a singular light
front. 

In view of this, it would be
desirable to find a formulation of classical electrodynamics that automatically
avoids any such incompatibilities. This is possible and in
Section~\ref{sec:conclusion} we discuss a whole class of such formulations 
having two representatives that are well-known since
the beginning of classical electrodynamics. But first, we end this section with
a short example that illustrates quantitatively 1) that the phenomenon of singular light
fronts can lead to significant radiation effects and 2) how the initial fields
encode the histories of their respective charge trajectories:\\

\emph{Quantitative example:} We reconsider the introductory example from
Section~\ref{sec:steps} of a charge, referred to as charge 1, having an initial
position and momentum $(\vv q_{1,0}=0,\vv p_{1,0})$ and an initial Coulomb field
$\vv f_{1,0}$, i.e., \eqref{eq:coulomb} Lorentz-boosted w.r.t.\ $\vv v(\vv
p_{1,0})$.  Whatever its future trajectory $(\vv q_1,\vv p_1)$ may be, its field
$\vv f_{1,t}$ will be of the form
\eqref{eq:coulomb_new}-\eqref{eq:coulomb_boundary} as depicted in
Figure~\ref{fig:coulomb}. In the following we will use SI units and charges smeared out by $\rho$
instead of point-charges, hence we regard $\vv F_{1,0}=\rho*\vv f_{1,0}$.
Let us suppose that the other initial fields
$\vv F_{j,0}$ of all $j=2,\dots,N$ other charges do not comprise free fields,
i.e., $\vv F^0_{j,0}=0$, and are such that charge 1 experiences a large initial
acceleration, say, $\vv a_{1,0}\sim (10^{17},0,0) \text m/ \text s^2$. Charge 2
is assumed to have initial position and momentum $(\vv q_{2,0},\vv p_{2,0})$ at
a sufficiently large distance, say, $\vv q_{2,0}=(0,10^2,0) \text m$, so that,
initially, it moves almost freely with velocity, say, $|\vv v(\vv p_{2,0})|\sim
10^4 \text m/\text s$.  Eventually, it will reach the vicinity of the light cone
of $(0,\vv q_{1,0}=0)$ smeared out by $\rho$.  Now, if charge 1 and 2 are made
out of clouds of, say, $Z=10^{13}$ electrons produced by electron guns and
collimated to balls of $\operatorname{diameter}(\rho)\sim 10^{-2} \text m$, the
acceleration of charge 2 is
\begin{equation*}
    \vv a_{2,t^*} \approx (\frac{e}{m} E_{1, t^*}^x ,0,0) \sim (-10^{14}
    ,0,0)\frac{\text m}{\text s^2},
\end{equation*} 
where $t^*$ is the time of arrival inside the smeared out light cone and
$E_{1, t^*}^x$ is the $x$-component of the electric field $\vv E_{1,t^*}$
computed with the help of \eqref{eq:coulomb_new} and \eqref{eq:lw}:
\begin{align*}
    E_{1, t^*}^x \approx \frac{eZ}{4 \pi \epsilon_0} \frac{-|\vv a_{1,0}|}{c^2
        |\vv q_{2,t^*}-\vv q_{1,0}| }.
\end{align*}
Note that, since for simplicity we assumed that the initial Coulomb field $\vv
f_{1,0}$ was already properly Lorentz-boosted, the distributions
\eqref{eq:coulomb_boundary} cancel in contrast to the introductory example in
Section~\ref{sec:steps}. In other words, (C1) is fulfilled which makes
\eqref{eq:coulomb_new} the only contribution to the field.  As a result of this
analysis, we find that within the time of traversal of the smeared out
light cone, $\Delta t\approx\operatorname{diameter}(\rho)/|\vv v_{2,0}|\sim
1\mu \text s$, there must be a sudden rise in emission of radiation of charge 2.
According to Lamor's formula, the resulting increase in power goes from almost zero to 
\begin{equation}
    \label{eq:P2}
    P_2= \frac{2}{3} \frac{Z^2 e^2 |\vv a_{2,t^*}|^2}{6 \pi \epsilon_0 c^3}
    \sim 1 W.
\end{equation}

One may now wonder, why the flank in radiation power increase is so steep. As
we discussed, the initial Lorentz-boosted Coulomb field of charge 1 encodes the
history of a charge with constant momentum $\vv p_{1,0}$. This tells the
Maxwell dynamics that there must be a sudden change in acceleration at time
$t=0$ from zero to $\vv a_{1,0}$ in order to fit the initial data, and
therefore, that a step in increase of radiation power (smeared out by $\rho$)
must be produced. In a more realistic scenario, however, charge 1 would first
have to acquire the initial acceleration $\vv a_{1,0}$ in the past $t<0$; to
match the above numbers, for instance, by entering a capacitor and falling
through a voltage of $U=10^4 \text V$ over a distance of $s=10^{-2} \text m$.
Depending on the duration of the acceleration process which may take
considerably longer than $\Delta t \sim 1\mu \text{s}$, a quite different
initial field $\vv F_{1,0}$ is produced. This time, it consists of the former
Lorentz-boosted Coulomb field plus a radiation part that was emitted during the
process of acceleration. In contrast to the above scenario, this additional
radiation will hit charge 2 much earlier before entering the light cone region,
and instead of generating a steep flank in radiation power of charge 2 from
zero to $P_2$ there will be a respectively smoother increase.  

In other words, and independently of the fictitious numbers we have used above,
this example shows that if the choice of initial values to the Maxwell-Lorentz
system \eqref{eq:lorentz}-\eqref{eq:constraints} imply a large initial
acceleration for charge 1, which may later on enforce charge 2 to generate
significant radiation \eqref{eq:P2}, the cause that led to this large initial
acceleration lies in the  history of charge 1 which is encoded in the initial
field $\vv F_{1,0}$. Hence, an unnatural choice for $\vv F_{1,0}$
such as the Lorentz-boosted Coulomb field above, which would imply that charge
1 did not accelerate for $t<0$ but then suddenly does at $t=0$, leads to the
peculiar effect of the steep flank in radiation power increase of charge 2.
Modelling more accurately how charge 1 acquired the initial acceleration in the
past will eliminate this effect.

This demonstrates the intimate connection between the history of a charge
trajectory and its generated field, which is the starting point of our
discussion of the formulation of classical electrodynamics in
Section~\ref{sec:conclusion}.

\section{Conclusion}
\label{sec:conclusion}

As demonstrated, in the case of point-charges,  the restriction of the solution
space of the Maxwell-Lorentz system \eqref{eq:lorentz}-\eqref{eq:constraints} to
smooth solutions does not allow a formulation in terms of an initial value
problem. Though a potential global solution is uniquely identified by its
initial data $(\vv q_{i,0},\vv p_{i,0},\vv f_{i,0})_{1\leq i\leq N}$, only very
special initial fields fulfilling the necessary condition (C2) lead to smooth
global solutions.  Furthermore, the information needed to restrict the initial
data according to (C2) would already require knowledge of the unknown solution.
Even for smooth charge distributions $\rho$, neglect in matching the initial
fields $\vv f_{i,0}$ to the history of the charge trajectories yields
rather arbitrary differences in the predictions as the example
in the preceding section illustrates. These circumstances suggest that we might
need to change the way we look at the solution theory for the Maxwell-Lorentz
system.

The starting point for such a consideration is the fact that the Maxwell field at one time
instant and the entire
trajectory of the charge that generated it are intimately intertwined beyond the
Maxwell constraint \eqref{eq:constraints}. This can be observed best when
imagining a single charge $i$ incoming from the remote past $t=-\infty$.
Considering, e.g., the case $\lambda=1$, any auxiliary trajectory
$(\tilde{\vv q}_i,\tilde{\vv p}_i)$ in the expression of the field $\vv
f_{i,t}$ in \eqref{eq:f-new}-\eqref{eq:f-free} is forgotten during a time
evolution from $t=-\infty$ to any finite time $t$ and so are any potential singular
fronts as they escape to spatial infinity with the speed of light.  Concerning
point-wise evaluation in any finite region of space-time, the Maxwell field in
\eqref{eq:f-new}-\eqref{eq:f-free} reduces to the expression
\begin{align}
    \label{eq:incoming-field}
    \vv f_{i,t} = \vv f_{i,t}^-[\vv q_i,\vv p_i] + \vv f^0_{i,t}.
\end{align}
Nothing changes in this argument and in the form of \eqref{eq:incoming-field} when the charge
trajectory $(\vv q_i,\vv p_i)$ is not prescribed but also develops
simultaneously to the evolution of the Maxwell fields, i.e., according to the fully coupled system
\eqref{eq:lorentz}-\eqref{eq:constraints}. Hence, stopping the dynamics at time
$t=0$ and starting it again in an initial value problem fashion dictates the
natural choice \eqref{eq:incoming-field} for the initial field at $t=0$.  This
means that the initial field $\vv f_{i,0}$ should be of the form \eqref{eq:f_0}
for a auxiliary trajectory $(\tilde{\vv q}_i,\tilde{\vv p}_i)$ that 
coincides with the actual one $(\vv q_i,\vv p_i)$ and that the free field $\vv
f^0_{i,0}$, as it evolves independently of the charges, equals the incoming free
field evolved from $t=-\infty$ to $t=0$. 

Hence, in the general case for any $\lambda\in[0,1]$, where also advanced
Li\'enard-Wiechert fields may occur, one would expect the Maxwell field to take
the form 
\begin{align}
    \label{eq:incoming-field-all}
    \vv f_{i,t} = \lambda\vv f_{i,t}^-[\vv q_i,\vv p_i]+(1-\lambda)\vv f_{i,t}^+[\vv q_i,\vv
    p_i] + \vv f^0_{i,t}.
\end{align}
Any compatibility condition, such as the Maxwell constraint
\eqref{eq:constraints}, (C1), and (C2), is now naturally fulfilled for all times
$t$. But this comes at a high price.  By \eqref{eq:incoming-field-all}, the
fields $\vv f_{i,0}$ at time $t=0$ depend on the entire history of the charge
trajectories which consequently means letting go of the initial value
formulation of classical electrodynamics.

In view of the above, however, such a step seems well grounded.
In Section~\ref{sec:ml-system}, it was already indicated when insisting on the
merely mathematical property of smoothness of solutions. But there, one might
even have been tempted to accept potential kinks in the charge trajectories,
say, as long as they decay fast enough.  However, the discussion above and in
Section~\ref{sec:admissible} shows that there is also a physical reason why the
initial value formulation is questionable, namely the fact that at each time
instant the entire history of a charge trajectory is already encoded in the
spatial dependence of its field. Therefore, when entertaining the thought that
charges are incoming from the remote past, the form of the Maxwell fields is
already presupposed by \eqref{eq:incoming-field-all} and the space of potential
solutions $(\vv q_i,\vv p_i,\vv f_i)_{1\leq i\leq N}$  of the Maxwell-Lorentz
system \eqref{eq:lorentz}-\eqref{eq:constraints} should consequently be
restricted to solutions having Maxwell fields $\vv f_{i,t}$ that fulfill
\eqref{eq:incoming-field-all}. 

Such a restriction is easily implemented in the fundamental equations of motion
\eqref{eq:lorentz}-\eqref{eq:constraints}. It simply means replacing the Maxwell
fields on the right-hand side of \eqref{eq:lorentz_b} with the explicit form given
in \eqref{eq:incoming-field-all}. This makes the Maxwell equations and
constraints \eqref{eq:maxwell}-\eqref{eq:constraints} redundant and turns the
coupled system of the ODEs \eqref{eq:lorentz} and PDEs
\eqref{eq:maxwell}-\eqref{eq:constraints}, only consisting of terms that are all
evaluated at the same time instant $t$, into the following system of ODEs that involve terms
depending on advanced or delayed times $t^\pm$ as given in
\eqref{eq:lw-abbreviations}:
\begin{align}
    \label{eq:delay}
    &\frac{d}{dt}\begin{pmatrix}
        \vv q_{i,t} \\ \vv p_{i,t}
    \end{pmatrix} 
    =  
    \begin{pmatrix}
        \vv v_{i,t}=\vv v(\vv p_{i,t})  \\
            \sum_{j =1}^N e_{ij} \vv L_{ij,t} \\
    \end{pmatrix},
    \\
    \vv L_{ij,t}&:=\int d^3x \, \rho( \vv x- \vv q_{i,t})
    [\vv E_{j,t} ( \vv x) + \vv v_{i,t} \wedge \vv B_{j,t}(\vv x)],
    \nonumber
    \\
    \vv F_{i,t}&=(\vv E_{i,t},\vv B_{i,t})
    \nonumber
    \\
    &=\rho*\Big(\lambda\vv f_{i,t}^-[\vv q_i,\vv p_i]+(1-\lambda)\vv f_{i,t}^+[\vv
    q_i,\vv p_i]\Big) + \vv F^0_{i,t}.
    \nonumber
\end{align}
Here, $\vv F^0_{i,t}$ denotes any given solution of the free Maxwell equations.  It
is interesting to note that  by virtue of \eqref{eq:kirchhoff}-\eqref{eq:f2} the
free fields $\vv F_{i,t}^0$, when prescribed in the remote past, are forgotten
should they have some spatial decay at spatial infinity \cite{KS00,BDD13}. In
this case, for $\lambda=1/2$, no self-interaction $e_{ij}=1-\delta_{ij}$ or $\vv L_{ii,t}=0$, and for point charges $\rho=\delta^3$, the
system of equations \eqref{eq:delay} is equivalent to the Fokker-Schwarzschild-Tetrode equations \cite{Fokker,Tetrode,Schwarzschild} as
used in Wheeler's and Feynman's investigation of classical radiation reaction 
\cite{WF45, WF49}.  They can be derived from a 
simple action principle \cite{Fokker,WF49}, and furthermore, allow a derivation of
Dirac's radiation damping term $\vv L^{ALD}_{ii,t}$ without the need of a mass renormalization procedure
\cite{WF45,BDDH13}.  Moreover, for $\lambda=1$, point charges $\rho=\delta^3$, and no initial free fields, the
resulting equations are equivalent to the Synge equations \cite{Syn40}.

The nature of these equations, involving a priori unbounded
state-dependent delays $t^\pm$, cf.~\eqref{eq:lw-abbreviations}, in the
definition of the Li\'enard-Wiechert fields \eqref{eq:lw}, renders a general
classification of solutions very difficult. In mathematics, this problem is
known as the \emph{electrodynamic N-body problem}. To this day, global
existence has only been established when considering two repelling charges and
restricting the motion of the charges to a straight line \cite{Dri77, Dri79,
Bau97, Ang90, HD15, BDDH16}. When constraining the charge trajectories at times $|t|\geq
T$, for arbitrary large but finite $T$, existence of solutions on $[-T,T]$ was shown for $N$
smoothly extended charges in three dimensions \cite{Dec10, BDD10}.
However, except for very special situations \cite{Dri79}, almost nothing is
known about uniqueness of solutions; see \cite{DDV12,BDDH13}.  It may turn out
that solutions can only be identified uniquely when whole stripes of trajectories
are specified.  Nevertheless, such types of state-dependent delay differential
equations are currently heavily under investigation in the contemporary
mathematics literature (see, e.g., \cite{Wal14} and the references therein) and
there is good reason to expect that their solution theory 
will soon be better understood.

\appendix

\section{Kirchhoff's formulas and explicit expression for the Maxwell fields}
\label{sec:appendix}
In this Appendix we explain how the solution formula
\eqref{eq:kirchhoff}-\eqref{eq:f2} for Maxwell's equations \eqref{eq:maxwell}-\eqref{eq:constraints}
can be derived from Kirchhoff's formulas. Afterwards we demonstrate the main
steps in the computation of the explicit expression
\eqref{eq:short-new}-\eqref{eq:short-free} from which
\eqref{eq:f-new}-\eqref{eq:f-free} and also
\eqref{eq:coulomb_new}-\eqref{eq:coulomb_boundary} can be inferred. Again we
omit the charge index and restrict ourselves to the point charge case $\rho =
\delta^3$ from which the corresponding result for general $\rho$ can be
inferred; all equalities are meant in distribution sense.

The Maxwell equations \eqref{eq:maxwell}, taking into account the constraint
\eqref{eq:constraints}, imply the following inhomogeneous wave equation
\begin{equation}
\label{eq:waveeq}
\square \vv f_t = 4\pi \begin{pmatrix} -\nabla & -\partial_t \\ 0 & \rot \end{pmatrix}\begin{pmatrix}\delta^3(\cdot - \vv q_{t}) \\ \vv v_{t} \delta^3(\cdot - \vv q_{i,t})\end{pmatrix} .
\end{equation}
Thanks to Kirchhoff's formulas \cite{Evans}, the unique solution 
$t\mapsto A_t$ to the 
initial value problem  $\square
A_t = 0, A_0:= A_t |_{t=0}, \dot{A}_0 := \partial_t A_t |_{t=0}$ 
can be expressed in the form of 
\begin{equation}
\label{eq:kirchhoff_0}
A_t = \partial_t K_t *A_0 + K_t *\dot{A}_0
\end{equation}
with $K_t$ as given in \eqref{eq:greensfn}.
Applying this formula to system \eqref{eq:waveeq} with initial values $\vv f_0$
and $\partial_t  \vv f_0 \mid_{t=0} = (\rot \vv b_{0} - 4 \pi \vv v_{0} \delta^3
(\cdot - \vv q_{0}) , - \rot \vv e_{0})$ at initial time $t_0$ we find an
expression for 
the unique solution given by
\begin{align}
\vv f_t &= \vv f_t^{(1)} + \vv f_t^{(2)},\label{eq:kirchhoff_F} \\
 \vv f_t^{(1)} &:= \begin{pmatrix}\partial_t & \rot \\ -\rot & \partial_t \end{pmatrix}K_{t-t_0} * \vv f_0   \label{eq:F1}\\
  \vv f_t^{(2)} &:= 4\pi \int_{t_0}^t ds \begin{pmatrix}
-\nabla & -\partial_t \\ 0 & \rot
\end{pmatrix}K_{t-s} * 
\begin{pmatrix}\delta^3(\cdot - \vv q_{s}) \\ \vv v_{s} \delta^3(\cdot - \vv q_{s})\end{pmatrix}. \label{eq:F2}
\end{align}
These formulas can also be found in \cite{KS00, spohn_dynamics_2008,Dec10}. Note, that in
general $\vv f_t^{(1)}$ and $\vv f_t^{(2)}$ do not solve Maxwell's equations
\eqref{eq:maxwell} individually. In the limit $t_0 \to \pm \infty$ and 
for $\vv f_0$ having some spatial decay
$\vv f_t^{(1)}$ vanishes and $\vv f_t^{(2)}$ coincides with the
advanced/retarded Li\'enard-Wiechert field $\vv f^\pm[\vv q,\vv p]=(\vv e^\pm, \vv b^\pm)$ as
given in \eqref{eq:lw}, which are solutions of the Maxwell equations
\eqref{eq:maxwell}-\eqref{eq:constraints}; see \cite{Dec10,BDD13}.

Next, we discuss the computation of \eqref{eq:short-new}-\eqref{eq:short-free}.
Without loss of generality we set $t_0 = 0$, and exemplary compute the term in
\eqref{eq:F2} coming from the upper left entry in the matrix of differential
operators evaluated at spacial point $\vv x \in \R^3$. The other two terms can be inferred analogously. We compute
\begin{align}
&4\pi \int_0^t ds\,(- \nabla_x K_{t-s} *\delta^3(\cdot - \vv q_s))(\vv x) \nonumber\\
&= -4\pi \int_0^t ds	 \frac{1}{4\pi (t-s)}\underset{\partial B_{|t-s|}(0)}{\int}d\sigma(y) \nabla_x \delta^3(\vv x-\vv y - \vv q_s)\nonumber\\
		&= - \int_0^{|t|} dr \frac{1}{r}\int_{\partial B_{|r|}(0)}d\sigma(y) \nabla_x \delta^3(\vv x-\vv y - \vv q_{t\pm r})\nonumber\\
        &=-\int_{B_{|t|}(0)}d^3y\frac{1}{\norm{\vv y}}\nabla_x \delta^3(\vv x-\vv y-
        \vv q_{t \pm \norm{\vv y}}). \label{eq:Aintegral}
\end{align}
In the next step, we employ the identity
\begin{equation*}
\nabla_x   \delta^3(\vv y-\vv y- \vv q_{t \pm \norm{\vv y}}) = L(\vv y)^{\pm}
    \cdot \nabla_y   \delta^3(\vv x -\vv y- \vv q_{t \pm \norm{\vv y}}),
\end{equation*}
for the matrix $L(\vv y)^{\pm}$ with entries
\begin{equation*}
L(\vv y)^{\pm}_{ij} := -\delta_{ij} \pm \frac{n_i v_j}{1\pm \vv n \cdot \vv
    v},
\end{equation*}
where the indices $i,j$ denote the components of the respective vectors and we
have used the abbreviations
\begin{equation*}
  \vv n := \vv n(\vv y) =\frac{\vv y}{|\vv y|}, \qquad
    \vv v := \vv v(\vv p_{t\pm |\vv y|}),
\end{equation*}
not to be confused with the notations $\vv n^{\pm}, \vv v^{\pm}$ in \eqref{eq:lw-abbreviations} and $\vv n_0, \vv v_0$ in \eqref{eq:abbreviations_0}.
For the $i$th component of the vector \eqref{eq:Aintegral} we then obtain
\begin{align}
    &(\ref{eq:Aintegral})_i= - \int_{B_{|t|}(0)}d^3 y \frac{1}{\norm{\vv y}}L(\vv
    y)^{\pm}_{ij} \partial_{y_j} \delta^3(\vv x -\vv y- \vv q_{t \pm \norm{\vv
    y}})\nonumber \\ & = \int_{B_{|t|}(0)}d^3 y\, \partial_{y_j}\left[
    \frac{1}{\norm{\vv y}}L(\vv y)^{\pm}_{ij}\right] \delta^3(\vv x -\vv y- \vv q_{t \pm
        \norm{\vv y}}) \label{eq:Ainside}\\ 
        &- \int_{\partial B_{|t|}(0)}d\sigma(y) \, n_j
        \frac{1}{\norm{\vv y}}L(\vv y)^{\pm}_{ij} \, \delta^3(\vv x -\vv y- \vv q_{t \pm
            \norm{\vv y}}) \label{eq:Aboundary}
\end{align}
by partial integration and Stokes theorem, where $\partial B_{|t|}(0)$ denotes the surfaces of the
ball of radius $|t|$ around the origin and $d\sigma$ the respective surface
measure. Furthermore, we adopt the Einstein summation convention regarding the
Latin indices. Next, it is convenient to carry out a change of variables according to
transformation
\begin{equation*}
    T(\vv y):= \vv y + \vv q_{t \pm \norm{\vv y}},
\end{equation*}
having Jacobi determinant $|DT(\vv x)|=1/(1 \pm \vv n^{\pm} \cdot \vv
v^{\pm})$.
Note that, as the trajectory $(\vv q,\vv p)$ is time-like, $T$ has an
inverse, and therefore, $T(B_{|t|}(0))=B_{|t|}(\vv q_0)$ holds.
After this transformation, \eqref{eq:Ainside} turns into
\begin{align*}
    &\underset{B_{|t|}(\vv q_0)}{\int}\hskip-.4cm d^3 z \, |DT(\vv z)|
\partial_{y_j}\left[ \frac{1}{\norm{\vv y}}L(\vv y)^{\pm} _{ij}\right] \bigg|_{\vv y=
    \vv z - \vv q^{\pm} }\delta^3(\vv x -\vv z)
\\
    &=
     \1_{B_{|t|}(\vv q_0)} \frac{1}{1\pm \vv n^{\pm} \cdot \vv v^{\pm}
}\partial_{y_j}\left[ \frac{1}{\norm{\vv y}}L(\vv y)^{\pm} _{ij}\right] 
\bigg|_{\vv y=
\vv x - \vv q^{\pm} },
\end{align*}
where we used the abbreviations introduced in \eqref{eq:lw-abbreviations}. The
derivatives can now be carried out in a straight-forward manner. 
The boundary term \eqref{eq:Aboundary} of the partial integration gives
\begin{align*}
& (\ref{eq:Aboundary}) = \int_{\partial B_{|t|}(0)} d \sigma(y) \frac{n_i }{(1\pm \vv n \cdot \vv v )\norm{\vv y}} \delta^3(\vv x - \vv y - \vv q_0)\\
&= \int_{\partial B_{|t|}(\vv q_0)} d \sigma(z) \frac{n_i(\vv z-\vv q_0)}{(1\pm \vv n(\vv z-\vv q_0) \cdot \vv v_0 )\norm{\vv z- \vv q_0}} \delta^3(\vv x - \vv z)\\
&= \delta(|t|- \norm{\vv x- \vv q_0}) \frac{n_{0,i}}{(1\pm \vv n_0 \cdot
\vv v_0) \norm{\vv x - \vv q_0}}.
\end{align*}
Carrying out the analogous computations as for \eqref{eq:Aintegral} and \eqref{eq:Aboundary} 
for the remaining two terms in \eqref{eq:F2} gives the following structure:
\begin{equation}
\label{eq:f2_explicit}
\vv f_t^{(2)} = \1_{B_{\Abs{t}}(\vv q_0)} \vv f_t^{-\sigma(t)}[\vv
q,\vv p] + \vv
    r_t^{-\sigma(t)}[\vv q_0,\vv p_0] ,
\end{equation}
where we have used the notation \eqref{eq:lw} and \eqref{eq:boundary}. The first
summand in \eqref{eq:f2_explicit} comprises all $d^3y$ integrals while the $d\sigma(y)$ integrals make up
the second summand.

Finally, in order to compute $\vv f^{(1)}_{t}$, given in 
\eqref{eq:F1}, we assume an initial field
$\vv f_0$ as in \eqref{eq:f_0} given in terms of a smooth auxiliary trajectory
$(\tilde{\vv q}, \tilde{\vv p})$, a homogeneous field $\vv f_0^0$,
and a parameter $\lambda \in [0,1]$. Plugging this choice for $\vv f_0$ into \eqref{eq:F1} gives
\begin{align}
\vv f_t^{(1)} &= \lambda \begin{pmatrix}\partial_t & \rot \\ -\rot & \partial_t \end{pmatrix}K_{t} * \vv f_0^-[\tilde{\vv q}, \tilde{\vv p}] \label{eq:Af1_1} \\
&+(1- \lambda) \begin{pmatrix}\partial_t & \rot \\ -\rot & \partial_t \end{pmatrix}K_{t} * \vv f_0^+[\tilde{\vv q}, \tilde{\vv p}]\label{eq:Af1_2} \\
&+ \begin{pmatrix}\partial_t & \rot \\ -\rot & \partial_t \end{pmatrix}K_{t}
*\vv f_0^0.
\label{eq:Af1_3}
\end{align}
Making use of the fact that the Li\'enard-Wiechert fields $\vv
f_t^{\pm}[\tilde{\vv q}, \tilde{\vv p}]$ solve Maxwell's equations and that $\vv
f_0^0$ is a homogeneous field, the three summands
\eqref{eq:Af1_1}-\eqref{eq:Af1_3} can be simplified 
according to
\begin{align*}
&\begin{pmatrix}\partial_t & \rot \\ -\rot & \partial_t \end{pmatrix}K_t * \vv f_{0}^{\pm}[\tilde{\vv q}, \tilde{\vv p}]  = \vv f_t^{\pm}[\tilde{\vv q}, \tilde{\vv p}] - \vv f_t^{(2)}[\tilde{\vv q}, \tilde{\vv p}]\\
&\begin{pmatrix}\partial_t & \rot \\ -\rot & \partial_t \end{pmatrix}K_t * \vv f_{0}^{0}  = \vv f_t^{0},
\end{align*}
where $ \vv f_t^{0}$ denotes the unique solution for the free Maxwell equations
to the initial value $\vv f_0^0$. Thanks to \eqref{eq:f2_explicit} we get
\begin{align*}
&\vv f_t^{(1)} = \lambda \left(  \vv f_t^{-}[\tilde{\vv q}, \tilde{\vv p}] - \vv r_t^{-\sigma(t)}[\tilde{\vv q}_0, \tilde{\vv p}_0] - \1_{B_{\Abs{t}}(\vv q_0)}\vv f_t^{-\sigma(t)}[\tilde{\vv q}, \tilde{\vv p}] \right) \\
&+ (1- \lambda) \left(  \vv f_t^{+}[\tilde{\vv q}, \tilde{\vv p}] - \vv r_t^{-\sigma(t)}[\tilde{\vv q}_0, \tilde{\vv p}_0] - \1_{B_{\Abs{t}}(\vv q_0)}\vv f_t^{-\sigma(t)}[\tilde{\vv q}, \tilde{\vv p}] \right)\\
&+ \vv f_t^{0}.
\end{align*}
Plugging this result together with \eqref{eq:f2_explicit} into
\eqref{eq:kirchhoff_F}, we arrive at the representation
\eqref{eq:short-new}-\eqref{eq:short-free} for the field $\vv f_t$. \linebreak

\textbf{Acknowledgments.}
The authors express their gratitude for the helpful discussions with
Detlef D\"urr, Michael Kiessling, Diego Noja, Herbert Spohn, and Roderich
Tumulka. This work was funded by the Elite Network of Bavaria through
the Junior Research Group `Interaction between Light and Matter'.

%\bibliographystyle{plain}
% \bibliographystyle{apsrev4-1}
% \bibliography{Literatur}

%

\end{document}